# Wide range and highly sensitive atomic magnetometry with Rb vapor

S. M. Iftiquar
Department of Physics, Indian Institute of Science, Bangalore – 560012, India
Email: smiftiquar@gmail.com

*Abstract*

*We have developed a technique in which Rb atomic response to weak magnetic field is high and an efficient rotation of linearly polarized laser beam results in efficient magnetometry. $^{85}Rb$ isotope has been used for the magnetometry in an ordinary vapor cell without any paraffin coating to its inner wall. A linear regime of Faraday rotation of about 25 $\mu T$ has been observed with atomic number density within the vapor cell of about $10^9$ $cm^{-3}$.*

**INTRODUCTION** .

High sensitivity low cost magnetometry is very essential in precise mapping of earth's magnetic field, magnetic resonance imaging (MRI) or magnetic pulse created by biological organs like heart brain etc. In this respect superconducting quantum interference devices (SQUID) have been used quiet extensively but because the SQUID devices are complicated to use and there are some external parameters like humidity that destroys high temperature SQUID [1]. A good alternative to this is vapor phase magnetometer [2] through the application of Faraday rotation.

Improvement of the Faraday rotation can be done in several ways, like introducing buffer gas within the vapor cell that increases lifetime of atomic state and excited atom do not decay faster due to collision with other atoms. Other technique is to use larger vapor cell and paraffin coating at the inner wall. Using such an advanced system a very sensitivity of magnetic field detection has been achieved with Cs vapor at its D1 transition line when laser frequency is modulated [3].

Another approach of improvement of sensitivity is pump-probe configuration. In this, a pump laser is passed normal to probe beam and the effect of rotation is observed in weak probe field [4]. In our system we use pump and probe beams in counter-propagating configuration and observe a significant enhancement of Faraday rotation.

**THEORY** .

In a Faraday configuration, an axial magnetic field is applied to a Rb cell and parallel to the magnetic field a linearly polarized probe beam is passed through the cell. This leads to a rotation of the polarization vector of the probe field. The polarization rotation depends upon atomic population in ground and excited state, preservation of atomic coherence with time and magnetic field intensity. What happens in this magnetometry is as follows: A linearly polarized laser beam can be considered as a composition of $\sigma^+$ and $\sigma^-$ laser light. In presence of magnetic field atomic hyperfine levels will be split up into Zeeman sub-levels. Atomic transition for $\sigma^+$ and $\sigma^-$ light will result in different refractive indices for the two circularly polarized components, $n^{(+)}$, $n^{(-)}$ respectively. So each component will face a different phase shift as it travels through the medium. At the exit from the cell the $\sigma^+$ and $\sigma^-$ light will acquire a relative phase shift, which while combined, will result in a rotation of the polarization vector. With the help of a polarizing cube beam splitter this polarization rotation is translated into probe light intensity variation. For a $90^o$ polarization rotation the intensity variation will be 100%, based on this the Faraday rotation angle can be expressed as

$$\phi = \frac{\pi . \delta I}{2I} \text{ radian}$$

where $\delta I$ is change in light intensity, I is total light intensity that is incident on photo-diode when probe polarization is $45^o$ to the vertical line. This rotation angle can be expressed as [5]

$$\phi = \frac{\omega}{2c}(n^+ - n^-)L$$

where $\omega$ is light frequency, c speed of light, L length of Rb vapor cell. The refractive indices $n^\pm$ can be derived in the following manner. The density matrix equation of motion of the atomic ensemble can be written as, ignoring power broadening

$$\frac{\partial \rho^{(1)}_{nm}}{\partial t} = -i\omega_{nm}\rho^{(1)}_{nm} - \frac{i}{\hbar}\left[H_I, \rho^{(0)}\right]_{nm} - \rho^{(1)}_{nm}\gamma_{nm}$$

Where m,n are the concerned energy levels where optical transition takes place, $\gamma_{mn}$ is decay constant for the first order perturbation of coherence $\rho_{mn}^{(1)}$, $\rho_{mn}^{(0)}$ is zeroth order perturbation of coherence matrix $\rho_{mn}$, $H_I$ is interaction Hamiltonian that is a product of time varying electric field of applied light (E(t)) and electric dipole moment of atom ($\mu$). The induced dipole moment will be [6]

$$P_{ij}^{(1)}(\omega) = \sum_{m,n} \frac{N(\rho^{(0)}_{mm} - \rho^{(0)}_{nn})}{\hbar} \frac{\mu^i_{nm}\mu^j_{mn}E_i(\omega)}{(\omega_{nm}-\omega)-i\gamma_{nm}}$$

i is direction of optical field polarization, j is direction of observed polarization, N is number density of atom in vapor cell, $\hbar$ is Plank's constant/$2\pi$ $\omega_{mn}$ is atomic transition frequency between the states m and n. In presence of applied magnetic field the atomic transition frequency will split by $g_F \mu_B B_z$, so

$$P_{ij}^{(1)}(\omega) = \sum_{m,n} A_{mn} \frac{\mu^i_{nm}\mu^j_{mn} E_i(\omega)}{(\omega_{nm} - (\omega \pm m_F g_F \mu_B B_z)) - i\gamma_{nm}}$$

where $A_{mn} = \frac{N(\rho^{(0)}_{mm} - \rho^{(0)}_{nn})}{\hbar}$, $m_F$ is Zeeman quantum number of hyperfine level. In order to take into account the thermal atoms that move about at a speed of about 270 m/sec, the above expression has to be averaged over Maxwell Boltzmann velocity distribution.

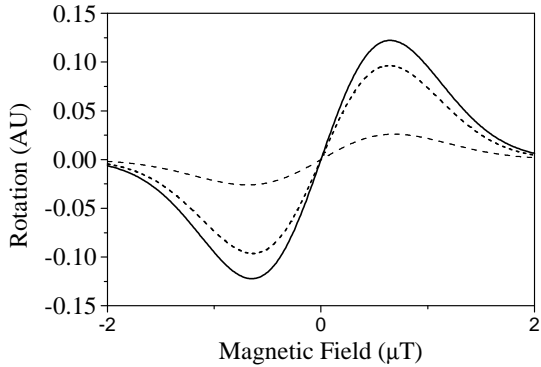

**Fig 1:** Effect of multiple Zeeman sublevels on Faraday rotation, dashed line is a contribution from $m_F$=+1, short dashed line is for $m_F$=+2 and the solid line is the result of both the Zeeman levels.

Figure 1 shows the effect of Maxwell Boltzmann speed distribution on Faraday rotation as well as effect of multiple Zeeman levels on the rotation angle and its slope. As visible in the figure, with the increase of number of Zeeman levels, the slope of the Faraday rotation increases. Thus the effect of the sensitivity increases with higher hyperfine quantum numbers.

**EXPERIMENTS**

The experimental system is shown in figure 2. The photodiode PD1 detects the Faraday rotation and the data is recorded by a digital oscilloscope. The Rb cell is of ordinary type, without any paraffin coating at its inner wall, and containing $^{85}$Rb (72.2%) and $^{87}$Rb (27.8%) isotopes at natural abundance. There is no buffer gas within the cell. Cell length is 5 cm and diameter 2.5 cm, beam diameter 2 mm, achieved by circular aperture. The cells are kept at room temperature and atomic number density of about $10^9$ cm$^{-3}$. The pump laser was locked to $^{85}$Rb F=3 → F'=(3,4) cross over.

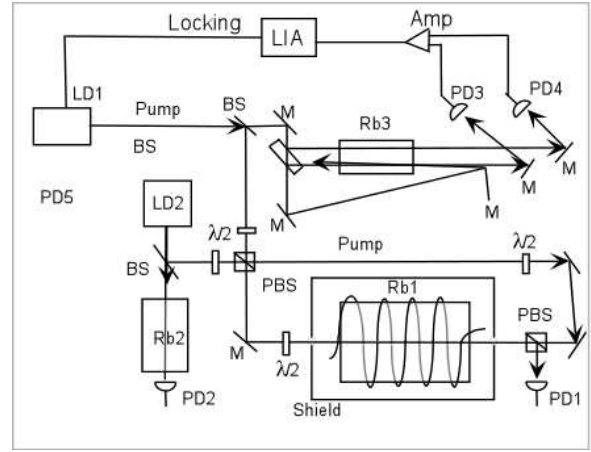

**Fig 2:** Schematics of experimental setup for magnetometric measurement in a counter-propagating pump-probe configuration LD1 pump laser, LD2 probe laser, LIA lock in amplifier, PD photodiode, Rb1, Rb2, Rb3 rubidium cells, Rb1 is for magnetometric measurement other cells are for saturated absorption spectra, 'Shield' is three layered magnetic shield for the Rb1, a solenoidal current carrying coil is wound over the Rb1 cell in order to create desired magnetic field within the Rb1..

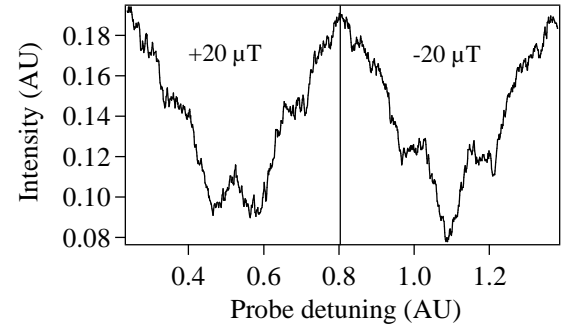

**Fig 3:** Probe spectra at different magnetic field condition, left part of the figure when +20 µT field is applied to the cell Rb1, while right hand part of the figure is for magnetic field of -20 µT.

**RESULTS AND DISCUSSIONS**

Figure 3 shows how Faraday rotation works in presence of the field. Looking at the two parts of the figure it is evident that the major change that takes place is near the F=3 to F'=(3,4) crossover region. The rest of the traces are characteristics to general absorption in vapor cell. The probe power is about 1.6 mW/cm$^2$, while the pump power is about 5 W/cm$^2$.

Figure 3 is taken under the condition of the probe polarization is at 45° to vertical so that PD1 signal increases for positive increase of magnetic field. This leads to increases and decreased PD1 signal for the two different magnetic fields, and the signal is most sensitive near F'=(3,4) cross over.

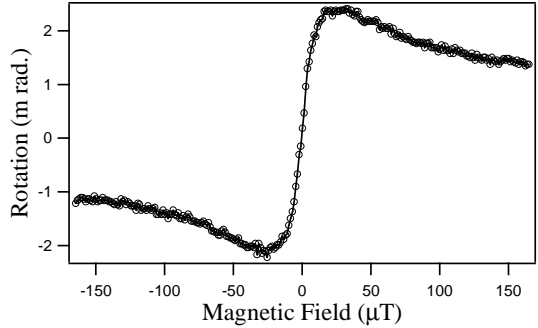

**Fig 4:** Experimentally observed Faraday Rotation with pump probe configuration..

Figure 4 shows experimental observation of non-linear Faraday rotation in a magnetic field range of $\pm 165\mu T$. A linear regime exists with a field span of ~$25\mu T$. Because of propagation direction of the pump-probe beams and their frequency, moving atoms that have velocity dependant Doppler shift of pump beam frequency resonating to $F = 3 \rightarrow F' = 4$, will be resonating to probe beam at $F = 3 \rightarrow F' = 3$. Although absolute magnitude of detection sensitivity less than other reported values [7] because of the type of Rb cell we used, yet in the technique much superior compared to single probe technique.

We have observed optical rotation with and without the pump beam and the rotation and sensitivity increases with pump beam intensity. Thus, this technique of magnetometry with saturation beam gives large enhancement of ordinary rotation as observed with single optical beam [7]. The enhancement factor is dependant upon pump laser intensity and is more than 1000 times than that without pump beam.

It is clear from the analysis that higher hyperfine quantum number is always better for magnetometry, that is why $^{85}$Rb atom is better sensitive to magnetic field than that of $^{87}$Rb atoms. It is well known that as one goes to very high sensitivity of magnetic field measurement the range of magnetic field that can be measured also reduces.

## CONCLUSIONS

Higher hyperfine is better for sensitive magnetometry and response per atom to magnetic field in our system is comparable to that of reference [7] because the sensitivity is ~$10^4$ times larger with atomic number density of ~$10^4$. In our setup a high sensitivity can be achieved without much reduction in range to linearity of Faraday rotation.

## ACKNOWLEDGEMENTS

In Acknowledgment, this work was supported by the Council of Scientific and Industrial Research of India.

## REFERENCES

1. P. Ripka ed , *Magnetic Sensors and Magnetometers*,., (Artech House, London, 2001).
2. Evgeniy B. Alexandrov, Marcis Auzinsh, Dmitry Budker, Derek F. Kimball and Simon M. Rochester, Valeriy V. Yashchuk,. J. Opt. Soc. Am. B, **22**, (2005) 7.
3. M. V. Balabas, D. Budker, J. Kitching, P. D. D. Schwindt, J. E. Stalnaker, J. Opt. Soc. Am. B, **23**, (2006) 1001.
4. S. Pustelny, D. F. Jackson Kimball, S. M. Rochester, V. V. Yashchuk, W. Gawlik,1 and D. Budker, Phys. Rev. A **73**, (2006) 023817.
5. X. Yang, Y. Chen, P. Cai, H. Wang, J. Chen, and C. Xia, Appl. Opt. **37**, (1998), p-4806.
6. S. M. Iftiquar, Conference Optical Society of India, 2007, paper number OSI_XXXIIIP-06.
7. D. Budker, D. F. Kimball, S. M. Rochester, V. V. Yashchuk, and M. Zolotorev, Phys. Rev. A **62**, (2000) 043403.